\documentstyle[preprint,aps,a4,12pt]{revtex}
\begin{document}
\draft
\hfill\vbox{\baselineskip14pt
            \hbox{\bf KEK-TH-511}
            \hbox{(\today)}}
\baselineskip20pt
\vskip 0.5cm 
\begin{center}
{\Large\bf Cylindrical versus Spherical Resonant Antennas \\
for Gravitational Wave Detection }
\end{center} 
\vskip 0.2cm 
\begin{center}
\large S. Alam$^{1,2}$\footnote{sher@theory.kek.jp}, S.~Haider$^{3}$
 ~ and ~Aurangzeb$^{3}$ 
\end{center}
\begin{center}
1: {\it Theory Group, KEK, Tsukuba, Ibaraki 305, Japan }\\
2: {\it Department of Physics, University of Peshawar, 
Peshawar, NWFP, Pakistan}\\
3: {\it Department of Physics, Islamia College, 
Peshawar, NWFP, Pakistan}\\
\end{center}
\vskip 0.2cm 
\begin{center} 
\large Abstract
\end{center}
\begin{center}
\begin{minipage}{14cm}
\baselineskip=18pt
\noindent
The principles and detection of gravitational waves by resonant
antennas are briefly discussed. But the main purpose of this short 
note is to compare the two geometries of resonant antennas, the 
well-known cylindrical to the spherical type. Some features of a 
two sphere observatory are also discussed.
\end{minipage}
\end{center}
\vfill
\baselineskip=20pt
\normalsize
\newpage
\setcounter{page}{2}
\section{Introduction}
 	The demonstration by Hertz of the existence of electromagnetic 
[EM] waves predicted by the Maxwell's equations was indeed a remarkable 
discovery. Einstein's equations of general relativity written some 
eighty years ago also lead us to a wave equation for gravitational 
waves [GW]. Despite bold pioneering efforts of eminent scientist 
Weber \cite{web61}and several 
ongoing international collaborations of excellent scientists \cite{coc95} 
we still await the direct detection of GW waves. The reasons are well-known
\begin{enumerate}
\item{} The gravitational coupling constant is approximately $10^{-38}$
times weaker than the electromagnetic coupling. 
\item{} The elementary mode of vibration induced by gravitational wave 
is {\em quadrupole} whereas an elementary mode excited by EM wave is 
{\em dipole}. A simple way to see why dipole radiation does not 
exist for gravity is because there is no negative mass and hence no 
gravitational dipole.
\end{enumerate} 
In the weak field approximation to Einstein's equations we can
treat GW waves as gravitons [spin two massless particle] propagating 
on {\em almost} flat space [i.e. very close to Minkowski space]. 
However the existence of gravitons should not be confused with the 
weak field limit. Their existence is implied by the radiative 
solutions of the Einstein's equations. Like EM waves GW are transverse 
and have two states of polarization. GW waves are more complicated
than EM waves because they unlike the latter self interact [or in
other words contribute to their own source].   

	The main goal is to first detect the GW waves directly. Once this
is done then one can use the gravitational radiation as a powerful
new probe and address questions, such as,
\begin{enumerate}
\item{}Examining interesting astrophysical systems like coalescing 
neutron-star binaries, black holes, supernovae, pulsars all of which 
are sources of gravitational radiation. 
\item{} GW from the early universe is expected to hold clues about
inflation. In particular GW excited during inflation as quantum
mechanical fluctuations is a key test of inflation and also allows
one to learn about specifics of the inflationary model\cite{tur96} 
\item{}Experimental information connected with string theory
[which has the promise of providing a unified theory of particles
and their interactions] may be obtained. GW detection would
help test general relativity against other competing theories
of gravitation such as scalar-tensor theories\cite{alam84}.
\end{enumerate}

	Detectors for GW waves fall into two broad categories,
i: Resonant Detectors and ii: laser interferometric devices.
We are concerned here with Resonant Detectors in particular
comparing the two geometries cylindrical [bar] with spherical.
The practical cylindrical bar antenna has been with for about 
30 years whereas the suggestion for the practical implementation
of spherical geometry is recent\cite{alb95,piz95}. Making a 
spherical antenna requires
more sophisticated technology in several areas such as casting,
cooling, suspension, and transducer attachment. However it
has some added advantages over it cylindrical counterparts
for it provides:  
\begin{enumerate}
\item{}Enhanced sensitivity, an order of magnitude better
than the corresponding cylindrical geometry.
\item{}Omni directionality.
\item{}Multi mode measurement capabilities.
\end{enumerate}
	The suggestion for using a spherical detector for 
gravitational wave was made as early as 1971 by 
Forward \cite{for71} where
it was indicated that by a suitable positioning of a set
of transducers on the sphere one could determine the
{\em direction}, the {\em amplitude}, and the {\em polarization}
of the gravitational wave. A free sphere has five degenerate
quadrupole modes of vibration that will interact strongly
with a gravitational wave. Each free mode can act as a
separate antenna, oriented towards a separate polarization
or direction. Wagoner and Paik \cite{wag77} showed that
the angle-averaged energy absorption cross section of a 
spherical antenna is much larger [ by a factor of 60] 
than a cylindrical bar [length to diameter ratio of 4.2] 
with the same quadrupole mode frequency.
So the question arises why these results were ignored?
One [main] reason is that a simple spherical detector is not
a {\em practical} detector. For one requirement of practicality
is a set of secondary mechanical resonators, which act as
mechanical impedance transformers between the primary vibrational 
modes of the antenna and the actual motion sensors, producing
the essential increase in the electromechanical coupling.
All successful cryogenic bar-type detectors have such
resonators. This practical consideration led Johnson and
Merkowitz \cite{joh93-95} to propose a method of positioning six radial
transducers on a truncated icosahedron to construct a nearly
spherical detector. They showed that a spherical detector cooled
to ultralow temperature can have sensitivity comparable or
even better than the first generation Laser Interferometric
Gravitational Wave Observatory [LIGO] detectors in the frequency
range around 1 kHz. A {\em network} of six cylindrical detectors
with appropriate orientation can cover the whole sky isotropically
[like a spherical detector] and have source direction resolution
\cite{cer93,zho95}. Such a network of six colocated cylindrical
detectors have has a sensitivity $\frac{1}{7}$ that of a
single spherical detector made of the same material and with the
same resonant frequency \cite{zho95}

\section{Resonant Antennas-Spherical and Cylindrical}
In the weak field approximation we can write
\begin{equation}
g_{\mu\nu}=\eta_{\mu\nu}+h_{\mu\nu}
\label{e1}
\end{equation}
where $h_{\mu\nu}$ [$h_{\mu\nu}\ll 1$] is the GW perturbation 
to the flat space time metric $\eta_{\mu\nu}$. As is known
and discussed in previous section GW waves have two states
of polarizations [$h_{+}$ and $h_{\times}$ with {\em quadrupole}
patterns. The perturbation $h$ will cause a change $\Delta L$
of a length $L$ between two free masses, such that
\begin{equation}
\Delta L=\frac{hL}{2}
\label{e2}
\end{equation}
assuming the wavelength of GW waves is much larger than
the mass separation L. 

It is well-known that the metric perturbation $h$ caused 
at the detector location by a GW burst generated due to 
conversion of mass $M_{gw}$
into gravitational radiation at a distance R from the 
detector position is,
\begin{equation}
h=\frac{1}{R\omega_{0}}\sqrt{\frac{8 G_{_{N}} M_{gw}}{c\tau_{gw}}},
\label{e3}
\end{equation}
where $\tau_{gw}$ is the duration of the GW burst, $G_{_{N}}$ is Newton's
constant and $\omega_{0}$ is the angular resonance frequency of
the antenna. It is instructive to write this in the form
\begin{equation}
h \approx 1.76\;10^{-17}\times\frac{1}{\omega_{0}}\times 
\sqrt{\frac{1}{10\tau_{gw}}}
\times\frac{10 {\rm Mpc}}{R}\times\sqrt{\frac{M_{gw}}{M_{s}}},
\label{e4}
\end{equation}
Assuming that $M_{gw}\sim 10^-2 M_s $, $\tau_{gw}\sim 10^-3$
and $\omega_0=2\pi\;1000$ rad/s in the above we find 
$h\approx 3\times 10^{-18}$ [for an event at the center of our 
galaxy, $R=8.5$ kpc] and $h\approx 2.6\times 10^{-21}$ [for
a source of GW waves located in the virgo cluster].

	The energy absorbed [captured] by the bar from GW burst
is related to the absorption cross section and the incident flux
[by the general definition of cross section] 
\begin{equation}
\sigma_{abs}=\frac{\Delta E_{abs}(\omega)}{\Phi(\omega)}
\end{equation}  
here $\Delta E_{abs}(\omega)$ is the energy absorbed by the
detector at frequency $\omega$ and $\Phi(\omega)$ is the
incident flux measured in ${\rm \frac{W}{m^2 Hz}}$. 
For a bar antenna [at the first longitudinal resonance mode]
\begin{equation}
\sigma_{abs}=\frac{8}{\pi}\frac{G_{_{N}}M v^2_s}{c^3}\sin^4\delta
\cos^2 2\Psi
\label{e5}
\end{equation}
where $v_s$ is the sound velocity, $M$ is mass of the antenna,
$\delta$ is the angle between the bar axis and direction of 
source and $\Psi$ is the angle between the plane [formed by
the bar axis and direction of source] and the polarization 
plane. The cross section increase with both $M$ and the
square of the sound velocity in medium $v_s$.
In the case of a spherical detector one finds for the absorption
cross section is
\begin{equation}
\sigma_{abs}=F_{ln}\frac{G_{_{N}}M v^2_s}{c^3}
\frac{\Gamma_{ln}}{(\omega-\omega_{ln})^2+\Gamma_{ln}^{2} /4}
\label{e6}
\end{equation}  
$F_{ln}$ is a dimensionless coefficient which is characteristic
of each mode. Assuming general relativity $F_{ln}$ is zero
unless $l=2$. $\omega_{ln}$ and $\Gamma_{ln}$ are respectively,
the mode resonance frequency and linewidth.

	To facilitate comparison between cylindrical and spherical
geometries we note $F_{21}\approx 3$ which is about 15\%
better than the cylinder's $8/\pi$ [Eq.~\ref{e5}]. If we
average over polarizations and directions the sphere's
cross section is a factor
$4.4$ better than the cylinder [assuming same masses for both] 
this result is in agreement with \cite{wag77,zho95,alb95} 
An aluminum [Al5056] cylinder [optimal orientation] of length=3 m, 
and diameter = 0.6 m has mass of 2.3 tons, the first longitudinal
mode has angular frequency of $2\pi \times 910$ rad/s
and the corresponding absorption cross section is 
$\sigma_{abs}=4.3\times 10^{-21}\;{\rm cm^2\;Hz}$.
A sphere [omnidirectional] of same material having
a diameter 3.1 m will have a mass of 42 tons. The
sphere has the same fundamental frequency. The absorption 
cross section is $\sigma_{abs}=9.2\times 10^{-20}\;{\rm cm^2\;Hz}$.
Similar numbers are noted in \cite{zho95,alb95}.
We note that although the length of the cylinder and
the diameter of the sphere are comparable the latter
is much more heavier, representing an advantage.
A conservative estimate of the sensitivity of 
this sphere is $h\approx 7\times 10^{-22}$. This
could be compared to a sensitivity of 
$h\approx 4\times 10^{-22}$ [in the same frequency range
i.e. $\sim 2\pi 1000$ Hz] for the best interferometers
and that is if GW arrives at optimum direction and
polarization.
 
Higher mode cross section values can be found by computing
$F_{22},\;\;F_{23},\;\;F_{24},\;\;F_{25}....$ \cite{alb95}.
One finds that we obtain larger cross sections at higher
harmonics for a spherical detector with respect to
its cylindrical counterpart. This suggests that by
using spherical detector as a xylophone we can scan
the frequency range $1-5$ kHz thus allowing us to study
the stochastic background of gravitational waves.
We note that a single sphere is not sufficient to identify
a GW event, for at least two antennas are necessary for minimum
coincidence analysis. A two sphere GW observatory has one 
important advantage over a network of several directional 
antennas with different orientations. The coincidence analysis 
is much simpler since the same amount of energy is absorbed 
by each detector while
for an array each member will absorb according to its orientation.
A coincidence analysis between spherical detectors is given
in \cite{zho95}. 
\subsection*{Present Detectors-Experimental Considerations}
	The vibrations in the bar are converted to electrical
signals [by electromechanical transducers], the electrical
signals are amplified and pass through filters to optimize
the signal to noise [SNR] ratio. The minimum vibration energy
which can be detected [SNR=1] is [from simple analysis]
\begin{equation}
E_{min}=k_B T_{eff}=\frac{k_B T}{\xi Q}+2 k_B T_n. 
\label{e7}
\end{equation}
T is the bar's temperature, $T_n$ is electronic noise 
temperature, $\xi$ is the fractional part of energy
available to the amplifier and Q is the quality factor
of all the apparatus. $T_{eff}$ is referred to as 
the effective noise temperature. It is clear that to 
compete against thermal noise [$k_B T$] in the bar $\xi Q$
must be as large as possible. Clearly $T$ and $T_n$
must be small. SQUID amplifiers can go down to the
quantum limits $T_n\sim 6\times 10^{-8}$ K but the
difficulty lies in matching them to the transducers.
For a FET amplifier the best value of $T_n$ is
more like $T_n \approx 0.1$ K. Let us see what
kind of numbers are involved if we assume a value
of $h\sim 3\times 10^{-21}$ which we obtained for 
the Virgo Cluster, Eq.~\ref{e4}. Now $h$ is related to
the mass of detector M, its length L, $\tau_{gw}$
and speed of sound in bar material by \cite{piz95}
\begin{equation} 
h=\frac{1}{\tau_{gw}}\times 
\sqrt{\frac{L}{v_s^2}}\times\sqrt{\frac{E}{M}}.
\label{e8}
\end{equation}
Using $h\sim 2.6\times 10^{-21}$ in the above equation
we find $E\sim 1.5\times 10^{-30}$ joule. Choosing this
value of $E$ as our $E_{min}$ in \ref{e7} we find that
the set $T\sim 40$ mK, $\xi Q \sim 10^6$ and $T_n \sim
.75 \times 10^{-7}$ satisfies equation \ref{e7}.
Meeting these requirements in an experiment [which is
conducted over long time] is quite demanding. From these
numbers we can easily appreciate the challenging task
of the experimentalists trying a direct detection of
gravitational waves.

	Let us now look at the characteristics of some the
resonant antennas [bar type] in world. We use the following
notation: Name of Experiment(Location, Detector mass in kg,
Temperature in Kelvin, Amplifier used, data taking, sensitivity h). 
Following this notation we have \\
CRAB(Tokyo,1200,4.2,parametric,1991,$2\times10^{-22}$[monochromatic])\\
EXPLORER(Rome,2300,2.5,SQUID,July 1990(1986),$7\times10^{-19}$)\\
NAUTILUS(Rome,2300,0.1,SQUID,1994-95,$3\times10^{-18}$)\\
ALTAIR(Rome,390,4.2,SQUID,-----,-----)\\
AURIGA(Legnaro,2300,0.1,SQUID,1995,-----)\\
ALLEGRO(Louisiana,2300,4.2,SQUID,June 1991,$7\times10^{-19}$)\\
NIOBE(Perth,1500,4.2,parametric,June 1993,$7\times10^{-19}$)\\
Moscow University(Moscow,1500,290,tunnel,1993,$7\times10^{-17}$).\\
For ALTAIR and AURIGA the sensitivity has not been listed for
they are yet to be fully tested. Sensitivity refers to a GW
burst lasting 1 ms, with the exception of the Tokyo group
which is searching for  the continuous GW from the CRAB pulsar.
The Rome antenna EXPLORER has reached sustained strain sensitivity
$h=6\times 10^{-19}$ for millisecond bursts over several consecutive
month period \cite{ast93}. ALLEGRO and the EXPLORER were also
operated in coincidence. They were aligned parallel to each
other and operated for 180 days [June 24th 1991- December 16th 1991]
so that the same gravitational wave burst would have produced
in the antennas signals with the same amplitude. Preliminary
analysis of the data gives a negative result \cite{piz95,coc95}.
In particular no coincident events were found 
above 200 mK \cite{coc95}.  
\section{Conclusions}
	In conclusion we have examined the advantages of the
spherical detector over its cylindrical counterpart and find
that our preliminary study confirms the optimistic results
reported in \cite{joh93-95,zho95,alb95}.  


\end{document}